\documentclass[preprint2]{aastex61}

\shorttitle{Flame-like EBs}
\shortauthors{Chen et al.}

\begin{document}

\title{Flame-like Ellerman Bombs and Their Connection to Solar UV Bursts}

\correspondingauthor{Hui Tian}
\email{huitian@pku.edu.cn}

\author{Yajie Chen}
\affil{School of Earth and Space Sciences, Peking University, Beijing 100871, China}

\author{Hui Tian}
\affiliation{School of Earth and Space Sciences, Peking University, Beijing 100871, China}

\author{Hardi Peter}
\affiliation{Max Planck Institute for Solar System Research, Justus-von-Liebig-Weg 3, 37077, G\"{o}ttingen, Germany}

\author{Tanmoy Samanta}
\affiliation{School of Earth and Space Sciences, Peking University, Beijing 100871, China}

\author{Vasyl Yurchyshyn}
\affiliation{Big Bear Solar Observatory, New Jersey Institute of Technology, 40386 North Shore Lane, Big Bear City, CA 92314-9672, USA}

\author{Haimin Wang}
\affiliation{Big Bear Solar Observatory, New Jersey Institute of Technology, 40386 North Shore Lane, Big Bear City, CA 92314-9672, USA}
\affiliation{Center for Solar-Terrestrial Research, New Jersey Institute of Technology, University Heights, Newark, NJ 07102-1982, USA}

\author{Wenda Cao}
\affiliation{Big Bear Solar Observatory, New Jersey Institute of Technology, 40386 North Shore Lane, Big Bear City, CA 92314-9672, USA}

\author{Linghua Wang}
\affiliation{School of Earth and Space Sciences, Peking University, Beijing 100871, China}

\author{Jiansen He}
\affiliation{School of Earth and Space Sciences, Peking University, Beijing 100871, China}

\begin{abstract}

Ellerman bombs (EBs) are small-scale intense brightenings in H$\alpha$ wing images, which are generally believed to be signatures of magnetic reconnection  around the temperature minimum region of the solar atmosphere. They have a flame-like morphology when observed near the solar limb. Recent observations from the Interface Region Imaging Spectrograph (IRIS) reveal another type of small-scale reconnection events, termed UV bursts, in the lower solar atmosphere. Though previous observations have shown a clear coincidence of some UV bursts and EBs, the exact relationship between these two phenomena is still under debate. We investigate the spatial and temporal relationship between flame-like EBs and UV bursts using joint near-limb observations between the 1.6--meter Goode Solar Telescope (GST) and IRIS. In total 161 EBs have been identified from the GST observations, and $\sim$20 of them reveal signatures of UV bursts in the IRIS images. Interestingly, we find that these UV bursts have a tendency to appear at the upper parts of their associated flame-like EBs. The intensity variations of most EB-related UV bursts and their corresponding EBs match well. Our results suggest that some of these UV bursts and EBs are likely formed at different heights during a common reconnection process.

\end{abstract}


\section{Introduction} \label{sec:intro}

Solar Ellerman bombs (EBs) are characterized as compact intense brightenings in images of the extended H$\alpha$ wings \citep{Ellerman1917, Ding1998, Georgoulis2002, Watanabe2008, Watanabe2011, Nelson2013, Nelson2015, Vissers2013, Vissers2015, Yang2013, Rezaei2015}. They reveal no obvious signatures in H$\alpha$ core images, and are generally believed to result from magnetic reconnection around the temperature minimum region (TMR). EBs are usually found in active regions (AR), though recent observations also show similar features in the quiet sun \citep{Rouppe2016, Nelson2017}. EBs often exhibit a flame-like morphology when observed near the solar limb with high-resolution instruments \citep{Hashimoto2010, Watanabe2011, Rouppe2016}. Recent three-dimensional (3D) magnetohydrodynamic (MHD) simulations have reproduced some properties of flame-like EBs \citep{Danilovic2017, Hansteen2017}. 

UV bursts are another type of small-scale reconnection events in the lower solar atmosphere, often observed in emerging ARs \citep[e.g.][]{Peter2014, Chitta2017, Zhao2017, Toriumi2017, Tian2018a, Young2018, Chen2019} and sunspot light bridges \citep{Toriumi2015,Tian2018b}. These events refer to intense compact brightenings in transition region (TR) images taken by the Interface Region Imaging Spectrograph \citep[IRIS,][]{DePontieu2014}. They were named ``hot explosions" or ``IRIS bombs" in some earlier studies \citep[e.g.][]{Peter2014,Tian2016}. The spectral profiles of some emission lines from the Si~{\sc{iv}} and C~{\sc{ii}} ions in UV bursts are significantly enhanced and broadened, often with several chromospheric absorption lines such as Ni~{\sc{ii}} 1335.20 and 1393.33 {\AA} superimposed.

Coordinated IRIS and ground-based observations show that some EBs are connected to UV bursts \citep{Kim2015, Vissers2015, Tian2016}, possibly indicating heating of the plasmas near the TMR to more than 10$^4$ K  \citep[e.g.,][]{Rutten2016}. 
However, 1D radiative transfer models of EBs suggest that the TMR cannot be heated to such high temperatures \citep{Fang2006, Bello2013, Berlicki2014, Hong2014, Fang2017, Hong2017a, Hong2017b,Reid2017}. On the other hand, MHD simulations of reconnection show that plasmas around the TMR can be heated to a few tens of thousand Kelvin if the magnetic field is strong and plasma $\beta$ is low \citep{Ni2016, Ni2018a, Ni2018b}. \citet{Hansteen2017} successfully reproduced EBs and UV bursts at different heights in 3D radiative MHD simulations. However, in these simulations EBs and UV bursts occur at different times and different locations, which is different from observations.

The exact relationship between EBs and UV bursts is still under debate. The fact that existing EB models cannot produce UV burst signatures possibly imply that these two types of events occur at different heights \citep[e.g.,][]{Fang2017}. Due to line-of-sight superposition, the possible height difference is difficult to be resolved from disk-center observations. However, near the limb EBs often reveal a flame-like structure due to the projection effect. With such a viewing angle, the possible height difference may be resolved from coordinated observations between IRIS and a large-aperture ground-based telescope. 

In this paper, we investigate the spatial and temporal relationship between flame-like EBs and their corresponding UV bursts using coordinated near-limb observations between IRIS and the 1.6--meter Goode Solar Telescope \citep[GST,][]{Cao2010}. Our results suggest that UV bursts and EBs likely occur at different heights during a common reconnection process.

\section{Observations and data reduction} \label{sec:obser}

We first searched coordinated near-limb observations between GST and IRIS satisfying the following criteria: (1) GST took images at H$\alpha$ core and $\pm$1 {\AA}; (2) IRIS took images with the 2832 {\AA} filter and at least one of the 1330/1400 {\AA} filters. Only two datasets meet our criteria. 

The first dataset was taken on 2017 May 27. IRIS performed a large coarse 16-step raster (120$^{\prime\prime}$ along the slit, 16 raster steps with a 2$^{\prime\prime}$ step size) of NOAA AR 12659 during 17:01--22:07 UT. The pointing coordinate was (740$^{\prime\prime}$,185$^{\prime\prime}$). Slit-jaw images (SJI) were taken with the 1330, 2796, and 2832 {\AA} filters, at cadences of 21 s, 21 s, and 83 s, respectively. We only used SJI 1330 and 2832 {\AA}, and degraded the temporal resolution of the 1330 {\AA} images to that of the 2832 {\AA} images. The spatial pixel size was $\sim$0.$^{\prime\prime}$17 for both SJI and spectral images. The spectral dispersion was  $\sim$0.051 {\AA}/$\sim$0.025 {\AA} in the near/far ultraviolet band. In this observation images of the Si~{\sc{iv}} 1393.76 {\AA} spectral window were not transmitted to the ground. The Visible Imaging Spectrometer (VIS) of GST took images at H$\alpha$ core, and H$\alpha$ wings at $\pm$1 {\AA}, $\pm$0.8 {\AA}, $\pm$0.6 {\AA}, $\pm$0.4 {\AA} and $\pm$0.2 {\AA} alternately during 16:43--22:43 UT, with a $\sim$53 s cadence at each wavelength position. The spatial pixel size of H$\alpha$ images was $\sim$0.$^{\prime\prime}$03.

The second dataset was obtained on 2015 June 25. IRIS performed a large sparse 16-step raster (120$^{\prime\prime}$ along the slit, 16 raster steps with a 1$^{\prime\prime}$ step size) of NOAA AR 12371 during 17:22--22:05 UT. The pointing coordinate was (656$^{\prime\prime}$,249$^{\prime\prime}$). Slit-jaw images were taken with all the four filters, each with a cadence of $\sim$17 s. Only the 1400 and 2832 {\AA} images were used. We did not use the 1330 {\AA} images, since the same UV bursts are present in both the 1400 and 1330 {\AA} images and UV bursts were identified primarily from the Si~{\sc{iv}} 1393.76/1402.77 {\AA} lines in many previous studies. The spatial pixel size was $\sim$0.$^{\prime\prime}$33 for SJI images. As the slit did not cross any EB-related UV bursts in this observation, the spectral data were not analyzed. The GST/VIS took images at H$\alpha$ core, and H$\alpha$ wings at $\pm$1 {\AA} and $\pm$0.6 {\AA} alternately during 16:49--21:46 UT. The cadence of H$\alpha$ images at each wavelength was $\sim$34 s, and the spatial pixel size was $\sim$0.$^{\prime\prime}$03. We degraded the temporal resolution of SJI images to that of VIS images.

The coalignment procedures between the IRIS and GST/VIS images for both datasets are similar: (1) We used the fiducial lines of IRIS to coalign the images taken in different SJI filters and spectral windows. (2) The IRIS SJI 2832 {\AA} images were internally aligned by using the cross-correlation technique to remove jitters. The obtained shifts were applied to the simultaneously taken 1330/1400 {\AA} and spectral images. (3) After filtering out bad frames, the GST/VIS images taken at each wavelength position were internally aligned using the cross-correlation technique. (4) By comparing commonly observed features such as sunspots and/or fibrils, VIS images taken at different wavelengths can be coaligned. (5) The coalignment between SJI and VIS images was finally achieved by comparing commonly observed features of sunspots and granules in the 2832 {\AA} and H$\alpha$ $\pm$1 {\AA} images frame by frame. 

\section{Results and discussion} \label{sec:discuss}

\begin{figure*} 
\centering {\includegraphics[width=\textwidth]{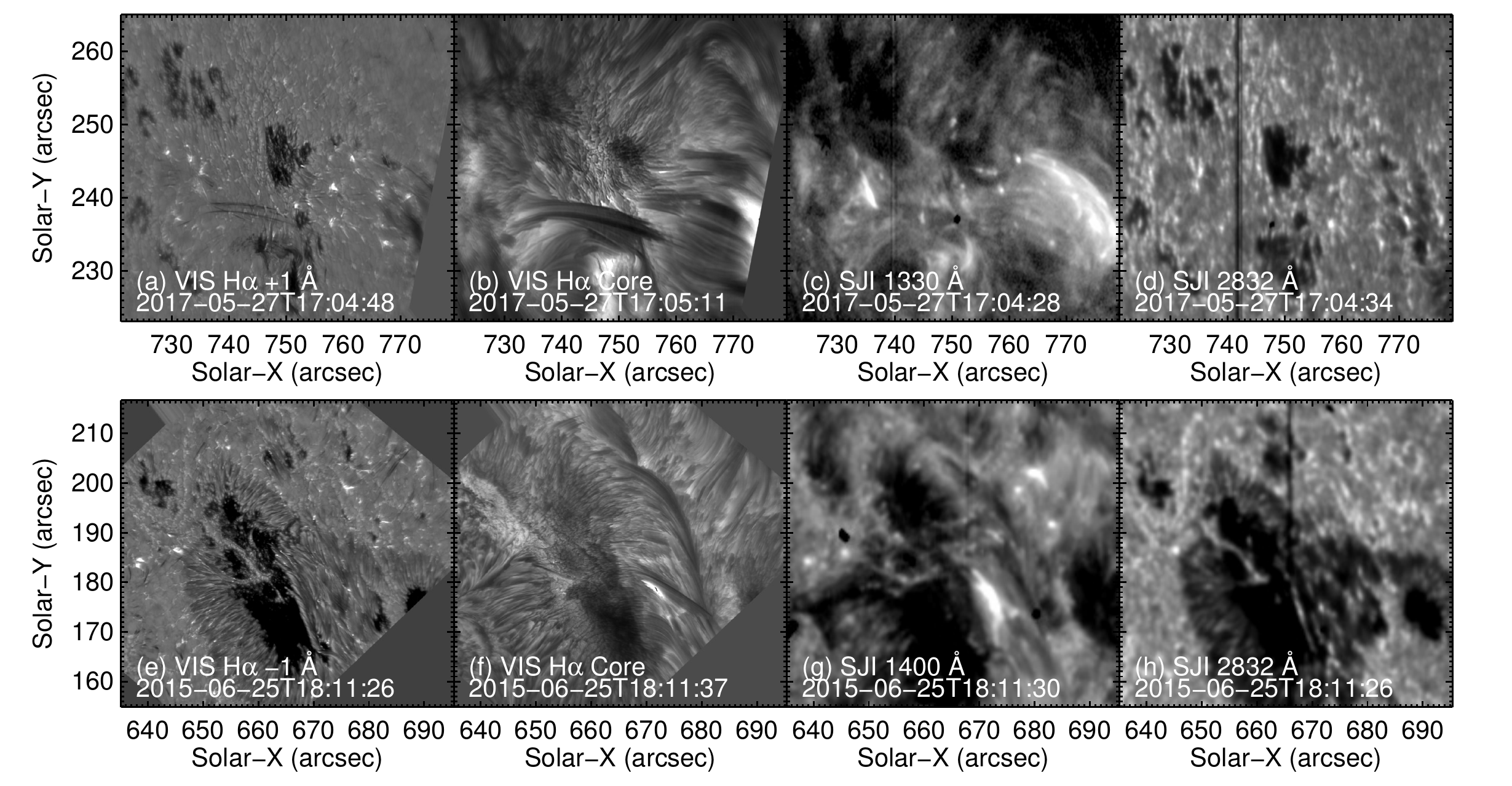}} 
\caption{Overview of the two observations. (a)--(d): GST/VIS H$\alpha$ $+$1 {\AA} and H$\alpha$ core images, IRIS/SJI 1330 and 2832 {\AA} images taken around 17:04:48 UT on 2017 May 27. (e)--(f): GST/VIS H$\alpha$ $-$1 {\AA} and H$\alpha$ core images, IRIS/SJI 1400 and 2832 {\AA} images taken around 18:11:26 UT on 2015 June 25. The intensities are shown in arbitrary units.} \label{f1}
\end{figure*}

Figure~\ref{f1} presents snapshots of the two observations. For each dataset only part of the field-of-view (FOV) is shown. From the H$\alpha$ $\pm$1 {\AA} image sequences, some compact brightenings can be identified around the sunspots. We used a criterion of 5$\sigma$ above the mean intensity at +1 {\AA} over the whole FOV to select EB candidates in the 2017 May 27 data. For the 2015 June 25 observation we used the --1 {\AA} images, since prominent dark flows in the +1 {\AA} images appear to obscure some EB candidates. For this dataset we lowered the threshold to 3$\sigma$, so that $\sim$0.15$\%$ pixels pass the thresholds for both datasets. We then examined the 11 or 5-point H$\alpha$ spectra of these EB candidates, and identified 161 EBs from the two datasets. In the meantime, we can see some transient compact brightenings from the IRIS 1330 and/or 1400 {\AA} images. Similarly, we used a threshold of 3$\sigma$ to identify UV bursts from the 1330 {\AA} images taken on 2017 May 27. For the 2015 June 25 observation, we used a threshold of 2.5$\sigma$ to identify UV bursts from the 1400 {\AA} images. A similar approach has been suggested by \cite{Young2018}. Twenty ($\sim$12\%) of the identified EBs appear to be associated with UV bursts. This fraction is not that different from the fractions found by \citet{Grubecka2016} and \citet{Tian2016}. 

\begin{figure*} 
\centering {\includegraphics[width=\textwidth]{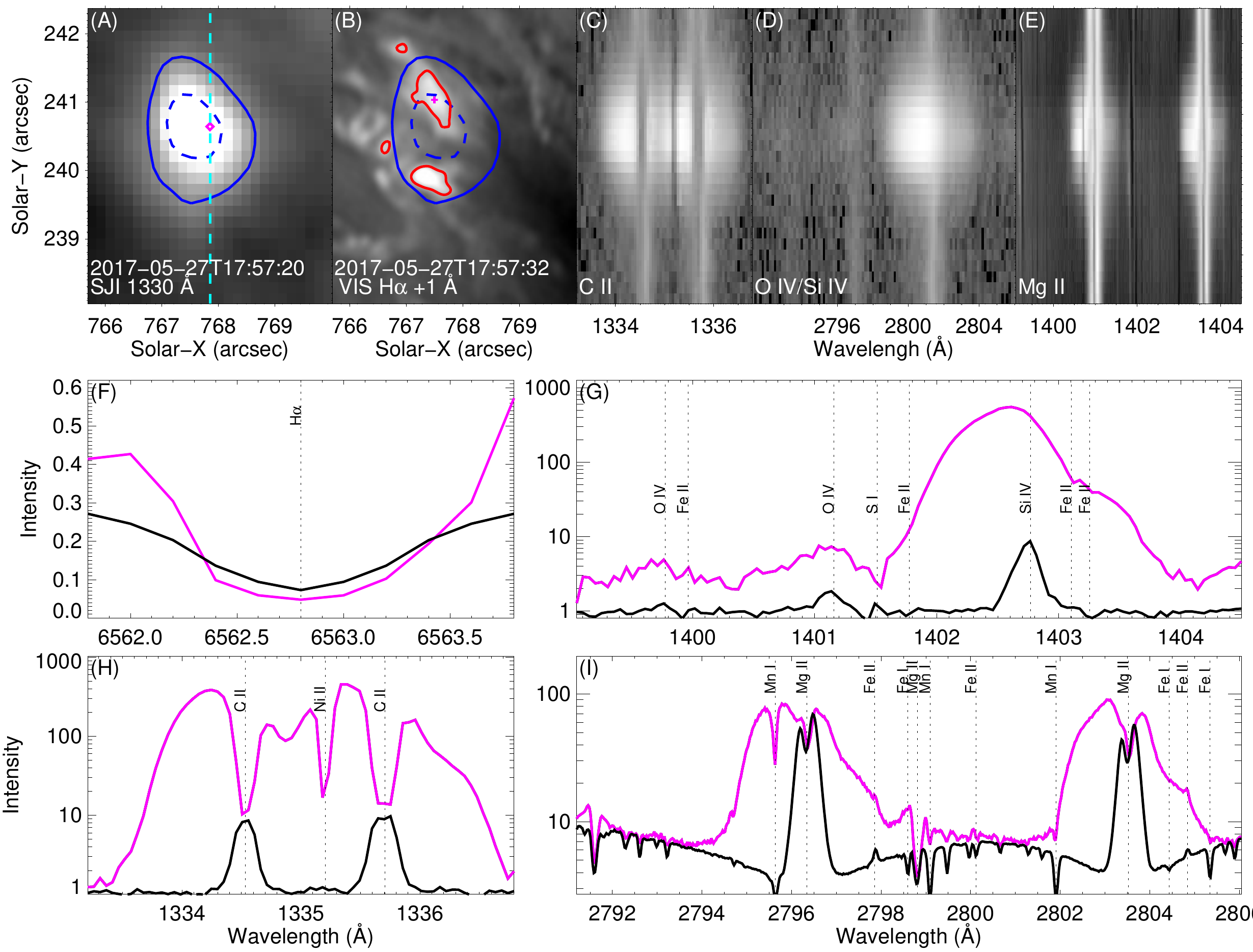}} 
\caption{IRIS and GST observations of an EB-related UV burst and its corresponding EBs. (A): IRIS/SJI 1330 {\AA} image taken at 17:57:20 UT on 2017 May 27. The cyan dashed line marks the IRIS slit position. (B): GST/VIS H$\alpha$ $+$1 {\AA} image taken at 17:57:32 UT. The solid and dashed blue contours in (A) and (B) indicate the location of the UV burst identified using a threshold of 3$\sigma$ and 5$\sigma$, respectively. The red contours in (B) mark the EB locations. (C)--(E): Simultaneously taken IRIS spectral images in three spectral windows. The intensities in (A)--(E) are shown in arbitrary units. (F): Normalized H$\alpha$ spectral profile (the purple line) at the location marked by the purple plus sign in (B). The black line represents the reference line profile. (G)--(I): IRIS spectra (purple lines, in the unit of countrate) at the location marked by the purple diamond in (A). The black lines represent the reference spectra from a plage region. The rest wavelengths of several spectral lines are marked by the vertical dashed lines in (F)--(I).
} \label{f2}
\end{figure*}

Note that a more strict identification of UV bursts relies on an examination of the IRIS spectra. However, only a few compact brightenings in the TR images were scanned by the IRIS slit. An example is presented in Figure~\ref{f2}, where the slit crosses a brightening in the IRIS/SJI 1330 {\AA} image taken around 17:57:20 UT on 2017 May 27. We examined the IRIS spectra at the brightening, and found that the Si~{\sc{iv}}, C~{\sc{ii}} and Mg~{\sc{ii}} line profiles are greatly enhanced and broadened. The absorption feature of Ni~{\sc{ii}} 1335.20 {\AA} is also obvious. Thus, this brightening is a typical UV burst \citep{Peter2014, Young2018}. We also notice the following characteristics for this UV burst: (1) the O~{\sc{iv}} 1401.156/1399.774 {\AA} lines are very weak; (2) the Mg~{\sc{ii}} k, h, and subordinate lines reveal a significant enhancement at the wings and no obvious enhancement in the cores; (3) the S~{\sc{i}} 1401.514 {\AA} line is broadened with a central reversal. These properties are similar to those of EB-related UV bursts \citep{Tian2016}. 
We can clearly see two EBs within the spatial range covered by this UV burst. The H$\alpha$ profile of the northern EB is presented in Figure~\ref{f2}(F), which shows an obvious enhancement at the wings and no enhancement at the core. The other EB has a similar H$\alpha$ profile. The UV burst appears to overlap more with the northern EB, which becomes evident if we increase the threshold from 3$\sigma$ to 5$\sigma$ for UV burst identification. In addition, compared to the southern EB, the intensity variation of the northern EB shows a much higher correlation with that of the UV burst, suggesting that the UV burst is likely associated with the northern EB only. Another possibility is that these two EBs form at the footpoints of a chromospheric loop, all of which is heated during reconnection \citep{Reid2015}. In this case the two EBs should be located in magnetic field regions with opposite polarities, which cannot be confirmed in this near-limb observation. It is worth mentioning that the reference H$\alpha$ line profile (averaged over the whole FOV) is asymmetric, and the intensities at the blue wing are lower than those of the red wing. This asymmetry is not physical. To remove this instrumental effect, we have forced the reference profile to be symmetric by multiplying the intensity at each wavelength position at the blue wing by a factor. The line profile at each EB was then scaled using these factors at multiple wavelengths. Note that this asymmetry does not affect our EB identification, since only the relative changes of spectral intensities need to be considered.



\begin{figure*} 
\centering {\includegraphics[width=\textwidth]{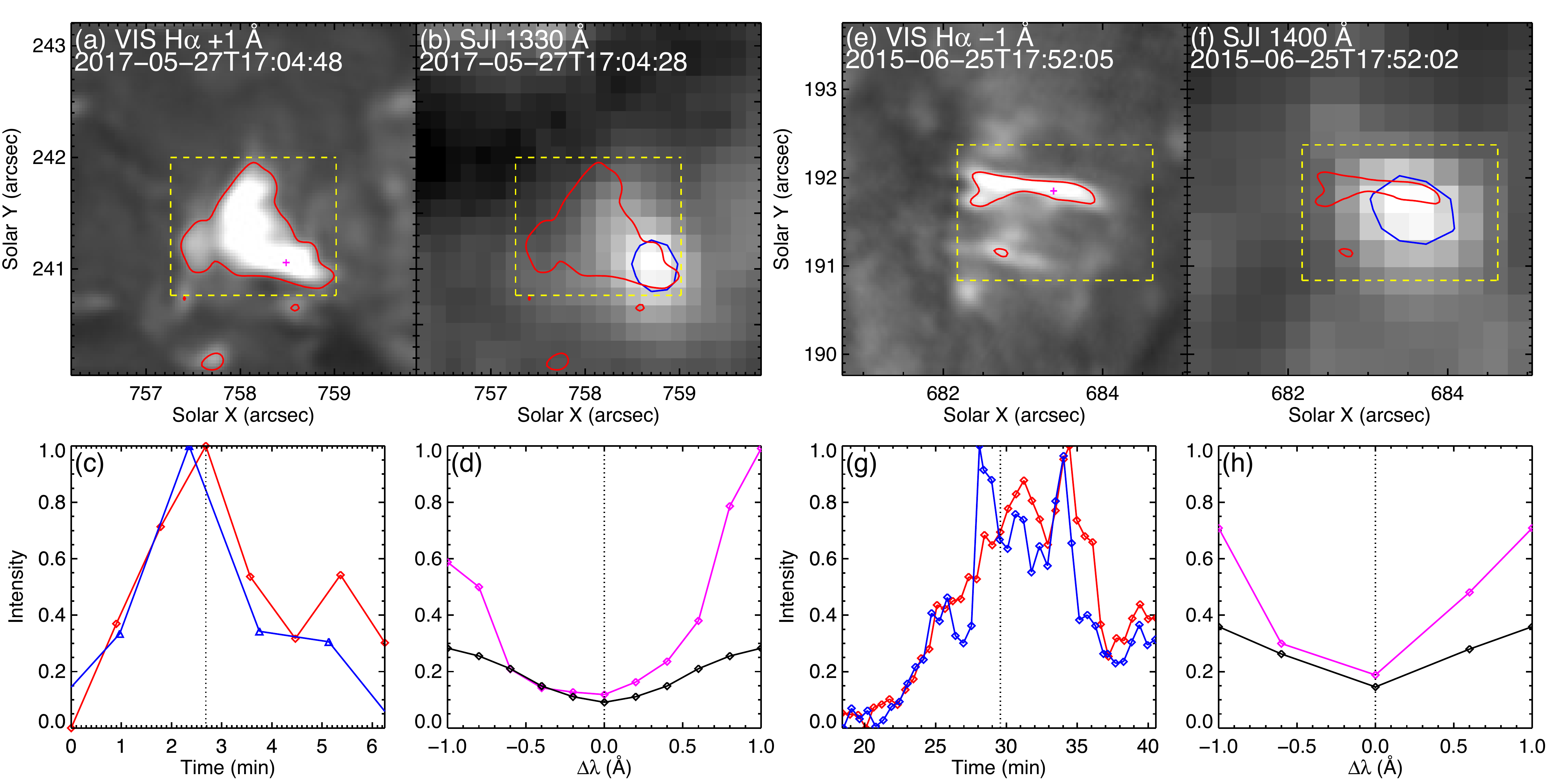}} 
\caption{Two examples of flame-like EBs and associated UV bursts. (a): GST/VIS H$\alpha$ $+$1 {\AA} image taken at 17:04:48 UT on 2017 May 27.  (b): IRIS/SJI 1330 {\AA} image taken at 17:04:28 UT. The blue contour marks the UV burst. The red contours in panels (a) and (b) outline the location of the EB. The intensities in (a) and (b) are shown in arbitrary units. (c) Temporal evolution of the total intensities of H$\alpha$ $+$1 {\AA} (red) and SJI 1330 {\AA} (blue) within the yellow box shown in (a) and (b). The vertical line indicates the time shown in (a). (d): The H$\alpha$ spectral profile (the purple line) at the location marked by the purple plus sign in (a). The reference profile is shown in black. The vertical line indicates the rest wavelength. The normalized intensities are shown in (c) and (d). (e)-(h): Same as (a)-(d) but for another event observed around 17:52:05 UT on 2015 June 25.} \label{f3}
\end{figure*}

Previous observations showed that most compact transient brightenings in SJI 1330/1400 {\AA} images have significantly enhanced and broadened Si~{\sc{iv}} and C~{\sc{ii}} line profiles \citep[e.g.,][]{Tian2016,Chen2019}. Thus, we treat all such brightenings that pass the thresholds mentioned above as UV bursts, though we could not examine the IRIS spectra of most brightenings due to the observational limitation. In total we have identified 61 UV bursts, and $\sim$31\% of them are connected to EBs. Figure~\ref{f3} presents another two examples. The first event was observed on 2017 May 27. The intensities at the extended wings of H$\alpha$ are greatly enhanced, while the intensities near the H$\alpha$ core do not change too much. Obviously, this is a typical EB. The EB exhibits a flame-like morphology in the H$\alpha$ wing image, which is typical in near-limb observations \citep{Hashimoto2010, Watanabe2011, Rouppe2016}. A visual inspection of the image sequence suggests that the UV burst is located at the top of the flame-like EB. Such a spatial offset likely reflects a height difference of the two phenomena. An earlier study by \citet{Vissers2015} also suggested that EBs are likely hotter in their tops. It is also interesting that the intensity variations of the EB and UV burst are very similar.

The second event presented in Figure~\ref{f3} was observed on 2015 June 25. Although we could only obtain 5-point H$\alpha$ spectra, we can still see obvious enhancements only at the wings and not at the core, consistent with the definition of an EB. The EB also exhibits a flame-like morphology. Moreover, it reveals an inverted ``Y''-shape structure at the footpoint, which is believed to result from magnetic reconnection between small magnetic bipoles and unipolar background fields \citep{Shibata2007}. Similar structures have also been identified from other high-resolution observations of EBs \citep{Watanabe2011,Tian2016,Tian2018b}. Similar to the first event, the associated UV burst is located at the top of the flame-like EB, and intensities of the EB and UV burst show correlated variations. 

\begin{figure}[ht!]
\plotone{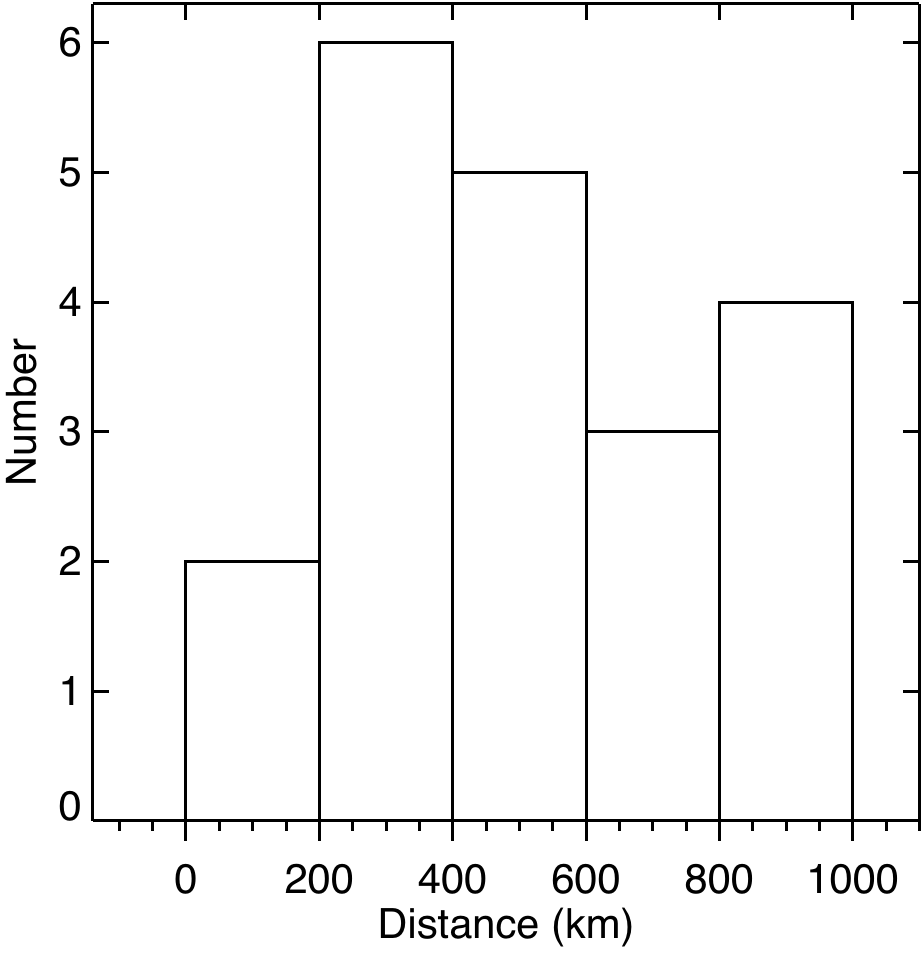} 
\caption{Histogram of the distance between the footpoint of an EB and the geometrical center of its associated UV burst.} \label{f4}
\end{figure}

We have examined all of the 20 identified EBs that are associated with UV bursts, and found that the UV bursts indeed have a clear tendency to appear at the upper parts of the EBs. To quantify the spatial offset between an EB and its associated UV burst, we measured the distance between the footpoint of the EB and the geometrical center of the UV burst. The projected spatial offsets are generally a few hundred km (Figure~\ref{f4}), suggesting that UV bursts are formed at least a few hundred km higher compared to their associated EBs. We found that all the EB-related UV bursts and their associated EBs exhibit an obvious partial overlap in their locations, possibly indicating a partial overlap of their formation heights. But we can not exclude the possibility that the overlap is caused by the projection effect.  

In addition, we found that the intensities of 15 EB-related UV bursts and their associated EBs reveal a similar temporal evolution (correlation coefficient $\geqslant$0.7). The coherent evolution strongly suggests that these EBs and their associated UV bursts are caused or modulated by a common physical process. Calculations of the EB intensities in the other five events are affected by unrelated brightenings or flows, resulting in low correlations between the intensities of EBs and UV bursts in these events.

\begin{figure*} 
\centering {\includegraphics[width=0.7\textwidth]{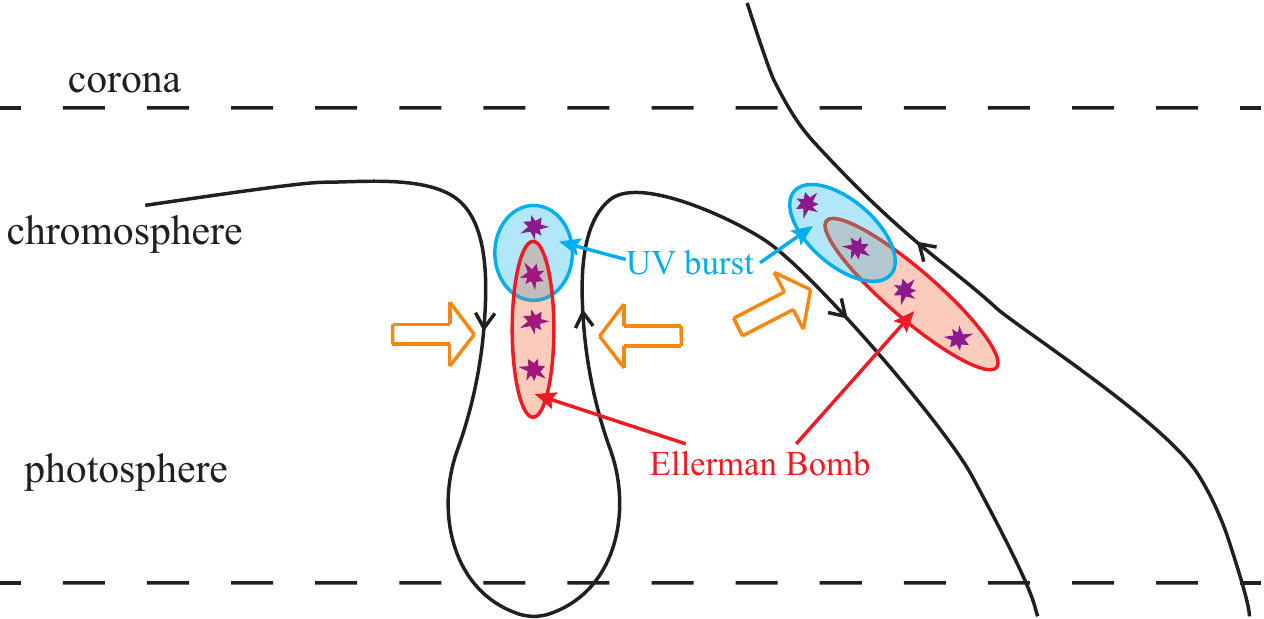}} 
\caption{A cartoon showing a possible connection between some EBs and UV bursts. The black solid lines, red stars and yellow arrows represent magnetic field lines, multiple X points within the current sheets and the reconnection inflows, respectively. } \label{f5}
\end{figure*}

It has been suggested that UV bursts may be produced by reconnection processes that are dominated by the plasmoid instability \citep[e.g.,][]{Innes2015,Ni2016}. In the 2.5D simulations of \citet{Nobrega2017} and \citet{Rouppe2017}, plasmoids are found in a wide range of atmospheric heights. In addition, recent models of \citet{Priest2018} and \citet{Syntelis2019} have shown that the plasma properties largely depend on the reconnection height. Inspired by our results and these models, we propose the following scenario (Figure~\ref{f5}) to explain the connection between some EBs and UV bursts: During flux emergence a nearly vertical current sheet may form and extend from the photosphere to the chromosphere. The current sheet could be located between the two sides of a U-loop produced through interaction between the emerging flux and convection, or at the interface between the emerging flux and an overlying field. As the current sheet becomes thinner, plasmoid instability is switched on. As a result, fast reconnection occurs and plasmoids are generated at different heights of the current sheet. Reconnections at chromospheric heights produce UV bursts, whereas lower-height reconnections within the same current sheet produce EBs. They may partially overlap in height. Such a scenario could explain the different temperatures of UV bursts and EBs, the spatial offset between them and their coherent intensity variations. Occasionally two EBs may occur close to each other simultaneously, as sketched in Figure~\ref{f5}. Then the associated UV bursts might merge to one larger event. This scenario would be consistent with the suggestion of \cite{Reid2015} and may explain the example shown in Figure~\ref{f2}. If the current sheet does not extend well above the TMR, we will only observe EBs. On the other hand, if reconnections occur only at greater heights, the resultant UV bursts will not be accompanied by EBs.

\section{Summary} \label{sec:sum}

Using two coordinated near-limb observations between IRIS and GST, we have investigated the relationship between UV bursts and EBs. We have identified 161 EBs from the GST observations, with most of them exhibiting a flame-like morphology. About twenty EBs show signatures of UV bursts in the IRIS TR images.

The UV bursts often appear at the higher parts of flame-like EBs, indicating that UV bursts are likely formed a few hundred km above the TMR. At such heights the plasma density is much lower than the photospheric density and thus is more likely to be heated to $\sim$8$\times$10$^4$ K. We have also found correlated variations of the intensities of most EBs and their associated UV bursts, suggesting a common cause or modulation of both phenomena. 

To explain our observational results, we propose that an EB and its associated UV burst are produced by magnetic reconnection processes at the photospheric and chromospheric heights, respectively, within a roughly vertical current sheet. 

\begin{acknowledgements}
This work is supported by NSFC grants 11825301, 11790304(11790300), 41574166, 41774183, and 41861134033, the Max Planck Partner Group program, AFOSR FA9550-15-1-0322 and NSF AST-1614457 grants. BBSO operation is supported by NJIT, NSF AGS-1821294 and NSFC-11729301 grants. The GST operation is partly supported by the Korea Astronomy and Space Science Institute, Seoul National University, and the Strategic Priority Research Program of CAS with Grant No. XDB09000000. IRIS is a NASA small explorer mission developed and operated by LMSAL with mission operations executed at NASA Ames Research center and major contributions to downlink communications funded by ESA and the Norwegian Space Centre. We thank Prof. Mingde Ding and Dr. Lei Ni for helpful discussions.

\end{acknowledgements}

\end{document}